# Complex Dirac-like Electronic Structure in Atomic Site Ordered Rh$_3$In$_{3.4}$Ge$_{3.6}$

Aikaterini Flessa Savvidou,[a,b] Judith K. Clark,[c] Hua Wang,[d] Kaya Wei,[a] Eun Sang Choi,[a,b] Shirin Mozaffari,[a] Xiaofeng Qian,[d] Michael Shatruk,[c] Luis Balicas,[a,b,]*

[a] *National High Magnetic Field Laboratory, 1800 E Paul Dirac Dr, Tallahassee, FL 32310, USA*
[b] *Department of Physics, Florida State University, 77 Chieftan Way, Tallahassee, FL 32306, USA*
[c] *Department of Chemistry & Biochemistry, Florida State University, 95 Chieftan Way, Tallahassee, FL 32306, USA*
[d] *Department of Materials Science and Engineering, Texas A&M University, College Station, TX 77843, USA*

Corresponding author: balicas@magnet.fsu.edu

**ABSTRACT.** We report the synthesis *via* an indium flux method of a novel single-crystalline compound Rh$_3$In$_{3.4}$Ge$_{3.6}$ that belongs to the cubic Ir$_3$Ge$_7$ structure type. In Rh$_3$In$_{3.4}$Ge$_{3.6}$, the In and Ge atoms choose to preferentially occupy, respectively, the 12$d$ and 16$f$ sites of the $Im\bar{3}m$ space group, thus creating a colored version of the Ir$_3$Ge$_7$ structure. Like the other compounds of the Ir$_3$Ge$_7$ family, Rh$_3$In$_{3.4}$Ge$_{3.6}$ shows potential as a thermoelectric displaying a relatively large power factor, $PF \sim 2$ mW/cmK$^2$, at a temperature $T \sim 225$ K albeit showing a modest figure of merit, $ZT = 8 \times 10^{-4}$, due to the lack of a finite band gap. These figures might improve through a use of chemical substitution strategies to achieve band gap opening. Remarkably, electronic band structure calculations reveal that this compound displays a complex Dirac-like electronic structure relatively close to the Fermi level. The electronic structure is composed of several Dirac type-I and type-II nodes, and even Dirac type-III nodes that result from the touching between a flat band and a linearly dispersing band. This rich Dirac-like electronic dispersion offers the possibility to observe Dirac type-III nodes and study their role in the physical properties of Rh$_3$In$_{3.4}$Ge$_{3.6}$ and related Ir$_3$Ge$_7$-type materials.

## INTRODUCTION

Intermetallic compounds exhibit a rich variety of compositions and crystal structures, offering a fertile playground for the discovery of new types of materials and physical phenomena. In recent years, studies of intermetallics with ever advancing theoretical and experimental methods have revealed a range of topological behaviors, understanding which relies on a thorough analysis of electronic band structures.

The low energy fermionic excitations of a wide range of materials, including high temperature $d$-wave superconductors,[1] topological insulators,[2,3] graphene,[4] and bulk semi-metallic systems like Cd$_3$As$_2$,[5,6] behave as massless Dirac particles instead of fermions that obey the Schrödinger Hamiltonian. These materials would display seemingly universal properties that would be a direct consequence of the linear or Dirac spectrum of their quasi-particles, such as a power-law temperature dependence for their fermionic contribution to the heat capacity.[7] Furthermore, the linear electronic dispersion around their Dirac or Weyl nodes and the nature of their electronic wave functions lead to remarkable electrical transport properties such as the observation of Klein tunneling, weak antilocalization, and unconventional Landau levels in graphene.[4,8,9]

Dirac type-I systems, such as graphene, display rotational symmetry around their Dirac quasi-momentum and display a point-like Fermi surface when the Fermi level $\varepsilon_F$ is precisely located at the node (Figure 1b). Dirac nodes can be classified according to the topography of the associated Fermi surfaces. For instance, if the parameters within the Dirac dispersion lead to a tilted Dirac cone with respect to $\varepsilon_F$, the cone can end up intersecting the Fermi level when the axial tilt surpasses a certain critical value (Figure

1d). Beyond this critical value, the Fermi surface evolves from a point to two crossing lines, leading to a finite density of states at the energy of the Dirac point, which is now classified as Dirac type-II.[10] However, at the critical tilting angle, between the Dirac type-I and type-II scenarios, the cone intersects $\varepsilon_F$ forming a Fermi surface composed of a single line characterized by a diverging density of states (Figure 1c). This kind of dispersion has been classified as Dirac type-III. It is characterized by very anisotropic effective masses, or localized states with heavy mass along one direction, i.e. along the intersecting line, that coexist with nearly massless states that propagate ballistically.

Type-III Dirac points have not yet been reported experimentally, while tilted type-I and type-II Dirac points have been reported for several compounds, e.g., $Cd_3As_2$[6] or $Na_3Bi$[11] for type-I and $PdTe_2$[12] for type-II. In most compounds, it is common for linearly dispersing bands to coexist with quadratically dispersing ones which tend to mask their overall physical properties. Available theoretical proposals to observe unconventional Dirac dispersions, such as Dirac type-III, have mostly relied on artificial photonic lattices that offer a great level of control over on-site energies and hopping elements.[13] The proposed schemes are based on the design of long-distance coupling of photons in lattices of resonators[14, 15] with the use of screw symmetries.[16] Thus far only $Zn_2In_2S_5$ was proposed as a possible candidate material to display Dirac type-III nodes in its dispersion.[17] These nodes would be relatively close to the Fermi level and, hence, capable of physically affecting ~~the~~ its charge carriers.[17] In black phosphorus, photoinduced Dirac type-I, type-II, and type-III nodes are predicted to become observable upon illumination with linearly polarized light.[18]

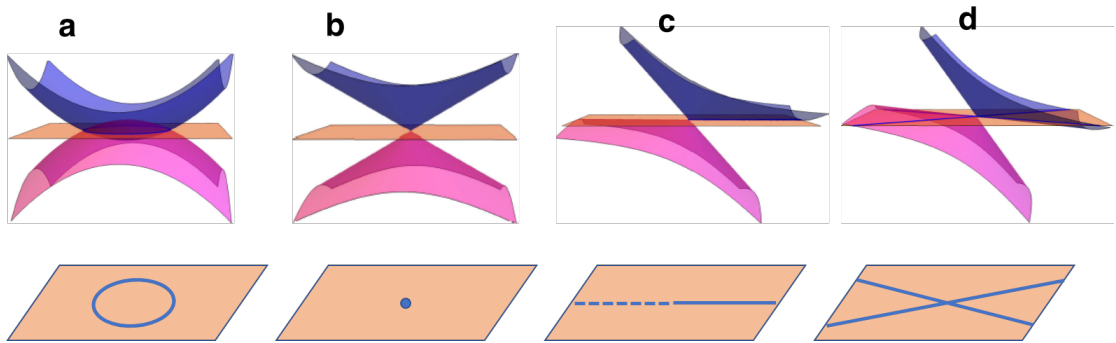

**Figure 1**. (a) Bands dispersing quadratically can intersect, leading to a nodal ring when this crossing coincides with the Fermi level $\varepsilon_F$ (depicted by the orange plane). (b) Linearly dispersing bands may meet at a point, forming a Dirac type-I node. (c) This Dirac dispersion can be tilted with respect to $\varepsilon_F$, leading, at a critical tilting angle, to a flat band at $\varepsilon_F$ (forming a Dirac line[17]), characterized by highly anisotropic effective masses. The Dirac line is associated to a non-trivial topological $Z_2$ invariant.[17] The associated Dirac node is classified as Dirac type-III. (d) When the tilting exceeds this critical value, conduction (blue) and valence (magenta) bands would intersect at $\varepsilon_F$, originating electron and hole Fermi surface pockets that touch at a Dirac type-II node.

Here, we report the synthesis of a novel cubic compound $Rh_3In_{3.4}Ge_{3.6}$ (nominally $Rh_3In_3Ge_4$), which crystallizes in the same space group as the well-known thermoelectric material, $Mo_3Sb_7$. This family of materials, which is based on the cubic $Ir_3Ge_7$-type structure (also known as a $Ru_3Sn_7$ type), was studied extensively in relation to its high-symmetry crystal structure and physical properties.[19-26] Depending on the elemental composition, they behave as metals or narrow band gap semiconductors. The latter type of behavior caused extensive studies of some of these compounds as promising thermoelectrics.[27] We demonstrate that $Rh_3In_{3.4}Ge_{3.6}$ is, to the best of our knowledge, the first well-characterized colored variant of the $Ir_3Ge_7$ structure type, i.e. the In atoms preferentially occupy one of the two crystallographically unique main-group metal sites. Remarkably, electronic band structure calculations reveal an incredibly rich Dirac structure in $Rh_3In_{3.4}Ge_{3.6}$, displaying Dirac type-I, Dirac type-II, and Dirac type-III nodes, all relatively close to the Fermi level. A Dirac type-III node is characterized by its intersection with a dispersionless band. Given that thermoelectric compounds, such as the tetradymites, have frequently been prime candidates for topologically non-trivial electronic structures,[28] we studied if this rich Dirac dispersion could lead to anomalous electrical or thermal transport properties. Although $Rh_3In_{3.4}Ge_{3.6}$ displays a respectable thermoelectric power factor relatively close to room temperature, its residual resistivity in the order of tens of μΩ cm translates into a modest carrier mobilities and low thermoelectric figure of merit (ZT). As is the case for graphene, we speculate that the off-stoichiometry of this compound introduces enough point-like defects/disorder to mask the electrical transport properties that one would expect for the stoichiometric compound. This type of disorder does not seem to significantly alter the material's phonon spectra and, therefore, its thermal conductivity to increase the value of ZT. Hence, we propose that improvements in stoichiometry might expose the presence of high mobility Dirac carriers, while chemical substitution to decrease metallicity or the synthesis of a microstructured material might improve the ZT. More importantly, $Rh_3In_{3.4}Ge_{3.6}$ offers a unique platform to explore the physical consequence of having multiple types of Dirac nodes relatively close to the Fermi level.

## RESULTS AND DISCUSSION

**Crystal Structure.** The crystal structure determination revealed that $Rh_3In_3Ge_4$ crystallizes in the cubic $Ir_3Ge_7$ structure type[19] (Figure 2). Compared to the parent binary structure, the In and Ge atoms form a nearly ordered arrangement by preferentially occupying Wyckoff sites 12*d* and 16*f*, respectively, in the cubic $Im\bar{3}m$ space group (Table 1). While the structure could be refined successfully with the complete separation of In and Ge over these atomic sites, the quality of the refinement improved substantially when mixed site occupancy was allowed. Nevertheless, the strong site preference was still observed, as the In/Ge site occupancy factors (SOFs) refined to 0.913/0.087 for site 12*d* and 0.165/0.835 for site 16*f*. The refined composition, $Rh_3In_{3.4}Ge_{3.6}$, is in much better agreement with the results of EDX analysis (Table 1), as compared to the idealized composition $Rh_3In_3Ge_4$ obtained for the structure without site mixing.

**Table 1**. Refined crystal structure parameters for $Rh_3In_{3.4}Ge_{3.6}$.

| Formula Unit: $Rh_3In_{3.40(5)}Ge_{3.60(5)}$ (EDX Analysis: $Rh_3In_{3.4(1)}Ge_{3.6(1)}$) | | | | | | |
|---|---|---|---|---|---|---|
| Space Group: $Im\bar{3}m$; Unit Cell: $a$ = 8.99999(5) Å, $Z$ = 4 | | | | | | |
| Atom | Wyckoff Site | $x$ | $y$ | $z$ | $U_{iso}$ | SOF |
| Rh1 | 12*e* | 0 | 0 | 0.33286(4) | 0.0112(1) | 1 |
| In1 | 12*d* | 0.25 | 0 | 0.5 | 0.0102(1) | 0.913(11) |
| Ge1 | 12*d* | 0.25 | 0 | 0.5 | 0.0102(1) | 0.087(11) |
| In2 | 16*f* | 0.16295(3) | 0.16295(3) | 0.16295(3) | 0.0122(2) | 0.165(8) |
| Ge2 | 16*f* | 0.16295(3) | 0.16295(3) | 0.16295(3) | 0.0122(2) | 0.835(8) |

Thus, $Rh_3In_{3.4}Ge_{3.6}$ should be considered as a colored variant of the $Ir_3Ge_7$ structure type. The parent structure has been discussed in detail in a number of papers,[20-22, 25, 29] and here we only emphasize the features introduced by the preferential distribution of the In and Ge atoms over the distinct crystallographic sites. Ignoring the slight disorder of In and Ge, each Rh atom is coordinated by 4 In atoms (at 2.7065(4) Å) and 4 Ge atoms (at 2.5770(6) Å), which form opposite faces of a square antiprism. Two such antiprisms share the In-based face to form dimers that protrude from the center of the unit cell toward the centers of its faces (Figure 2a). These dimers are joined through vertex-sharing Ge atoms, which thus form hollow cubes (Figure 2b). A rather short distance between the Rh atoms, 3.008(2) Å, is found across the shared face in the antiprism dimer. This short separation is contrasted with the much longer distance of 4.237(2) Å to the Rh atom of the adjacent dimer. The In–In distances in the shared face of the square antiprism are 3.18198(2) Å. This framework, built of the dimers of face-shared square antiprisms, is penetrated by an identical framework (Figure 2c), which leads to the observation of even shorter distances of 2.714(1) Å and 2.9331(6) Å between the In atoms of the two interpenetrating frameworks.

To the best of our knowledge, $Rh_3In_{3.4}Ge_{3.6}$ provides the first example of strong preferential site distribution of post-transition elements conclusively demonstrated by the direct crystal structure determination for any $Ir_3Ge_7$-type ternary compound containing main-group elements distinguishable by X-ray diffraction. Hulliger prepared compositions $Ir_3Ga_3Sn_4$ (X/Y = Ga/Sn, In/Ge), but only established their structure type and unit cell parameter.[29] The structure of $Rh_3In_5As_2$ was determined from powder X-ray diffraction data, but In and As were restricted to have the same distribution over both 12$d$ and 16$f$ sites.[30] The only previous examples where the SOF values were rigorously determined for a mixture of main-group elements in the 12$d$ and 16$f$ sites are series $M_3Sn_{7-x}Li_x$ and $M_3Sn_{7-x}Mg_x$ (M = Rh, Ir). In the former, the Li atoms showed a larger preference toward substituting Sn atoms in the 16$f$ site,[31] while an opposite trend was observed for the latter, where the Mg atoms preferentially entered the 12$d$ site.[23]

Häusermann *et al.* analyzed in detail the relationship between the electronic band structure and chemical bonding in these materials,[19] to demonstrate that the structure remains stable within a relatively narrow range of valence electron concentrations (VEC), i.e. 51-55 electrons per formula unit (f.u.). In the present compound, the VEC is equal to 51.6 electrons per f.u. for the experimentally determined composition, $Rh_3In_{3.4}Ge_{3.6}$, or 52 electrons per f.u. for the idealized composition, $Rh_3In_3Ge_4$, – well within the range of the electronic stability criterion suggested by the earlier theoretical analysis.

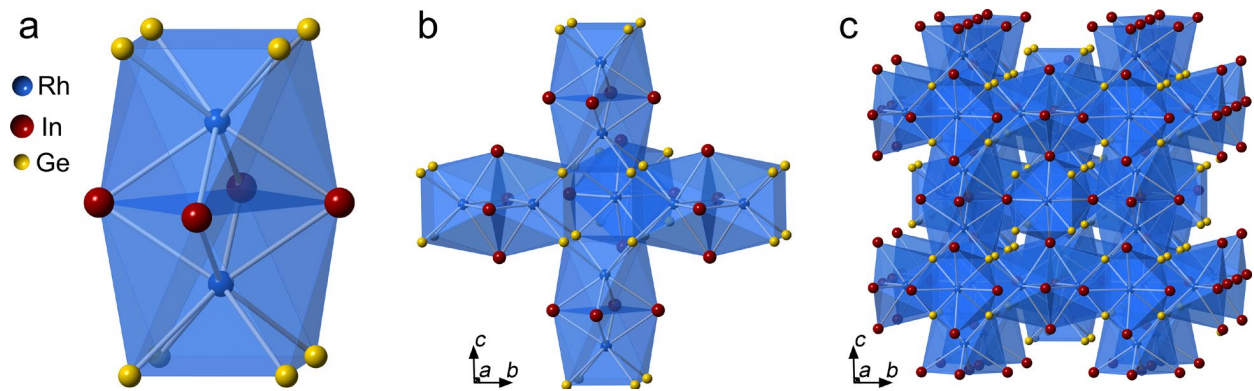

**Figure 2.** The crystal structure of $Rh_3In_3Ge_4$: face-sharing dimers of Rh-centered $In_4Ge_4$ antiprisms (a) are connected via edge-sharing (b) into a framework; two such frameworks interpenetrate to form the three-dimensional $Ir_3Ge_7$-type structure (c). The panel (b) also emphasizes the hollow $Ge_8$ cube created by face-sharing of the antiprism dimers.

**Physical Properties.** To characterize the physical properties of $Rh_3In_3Ge_4$, we performed electrical and thermal transport, as well as heat capacity and magnetic susceptibility measurements. The temperature-dependent resistivity, $\rho_{xx}$, measured through the conventional four terminal method on a polished $Rh_3In_3Ge_4$ sample (Figure 3a) reveals metallic behavior with a modest residual resistivity ratio, $\rho_{xx}(300K)/\rho_{xx}(1.8K)$ = 1.28 (Figure 3b). This small value is due to the relatively high residual resistivity $\rho_{xx}$ ($T \to 0$ K) = $\rho_o \cong 34$ µΩ cm. Below $T \sim 20$ K, $\rho_{xx}$ saturates at the value $\rho_o$, thus indicating that disorder scattering dominates its electrical transport properties at low temperatures. We extended these measurements down to $T = 0.1$ K to check for the possibility of superconductivity, as reported[32] for the isostructural compound $Mo_3Sb_7$, but none was observed.

Magnetic susceptibility ($\chi$) measurements revealed diamagnetic behavior over the entire temperature range (Figure 3d). The very small Curie-like tail observed for $T < 20$ K is not an indication for magnetic impurities, since $\chi^{-1}(T)$ does not display a linear dependence on $T$ even at the lowest temperatures. Most likely, it just reflects the evolution of the density of states at the Fermi level and its effect on Landau diamagnetism.

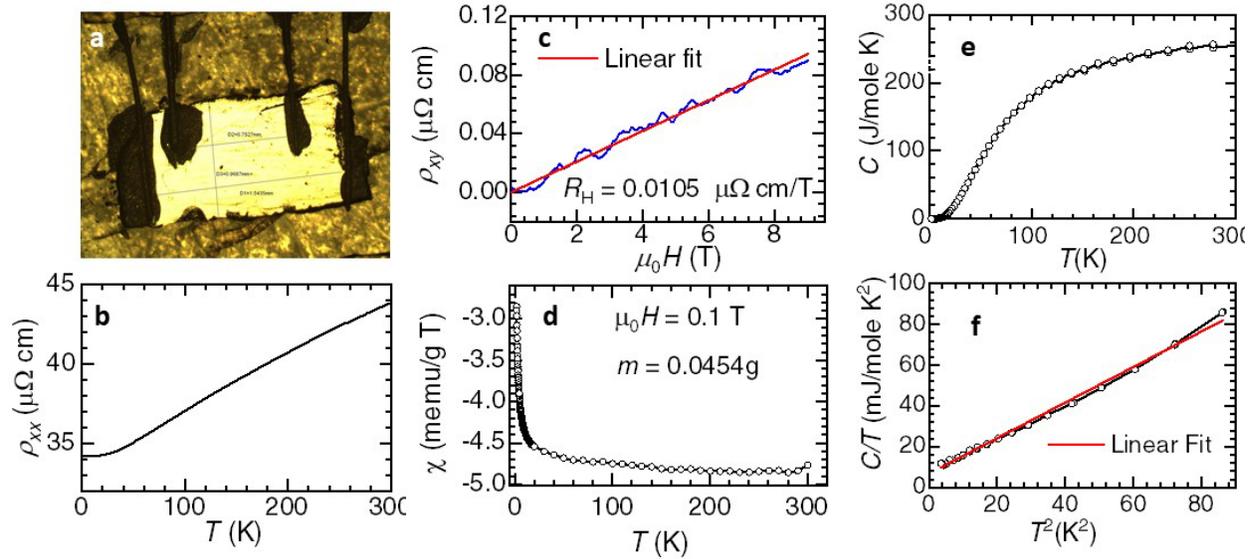

**Figure 3.** (a) Picture of a polished $Rh_3In_3Ge_4$ single crystal (0.154×0.0969×0.0015 cm³) with attached leads used for resistivity measurements. (b) Resistivity $\rho_{xx}$ as a function of temperature $T$. (c) Hall resistivity $\rho_{xy}$ as a function of the magnetic field $\mu_0H$ up to 9 T. The blue line depicts data points while the red line is a linear fit; $R_H$ is the extracted Hall coefficient. (d) Magnetic susceptibility $\chi$ as a function of $T$ under an applied field of $\mu_0H = 0.1$ T. (e) Heat capacity as a function of $T$. (f) Heat capacity normalized by $T$ as a function of the $T^2$ for $T \le 8$ K. Black markers correspond to experimental data; the red line corresponds to a linear fit.

To evaluate carrier density and their mobilities, which, as exemplified by graphene,[33] are expected to be rather high in systems dominated by Dirac-like quasiparticles, we performed Hall-effect measurements at temperatures as low as $T = 2$ K under magnetic fields up to $\mu_0H = 9$ T (Figure 3c). The Hall coefficient ($R_H$) was obtained from the linear fit (red line) to the data (blue line), yielding $R_H = 0.0105$ µΩ cm/T. The positive slope of the Hall resistance indicates that holes are the main charge carriers in this compound. From $R_H$ we extract a hole concentration, $n_H = 5.9\times10^{22}$ cm⁻³, and a hole mobility, $\mu_H = 3.07$ $\frac{cm^2}{V \cdot s}$. The modest $\mu_H$ value suggests that the off-stoichiometry of this compound and related crystallographic disorder dominate its transport properties. Notice that the value of the Fermi liquid coefficient, $A = 3.56\times10^{-4}$ µΩ cm K⁻² in $\rho_{xx} = \rho_0 + AT^2$, is relatively modest when compared to those of strongly correlated systems.[34] This observation indicates that electron-electron scattering is relatively modest in this compound.

Therefore, it is possible that the near-perfect $Rh_3In_3Ge_4$ stoichiometry might lead to considerably higher carrier mobilities.

Heat capacity ($C$) measurements (Figure 3e) indicate that $C$ follows the standard Debye $T^3$ law at low temperatures.[35] A linear fit was applied for the $C/T$ versus $T^2$ dependence (Figure 3f) to extract the electronic contribution, yielding a Sommerfeld coefficient $\gamma = 6.58$ mJ/mole K$^2$. This value is comparable to those of conventional metals[36] albeit the ratio $A/\gamma^2 = 0.8 \times 10^{-5}$ $\mu\Omega$ cm mol$^2$ K$^2$/J$^2$ is very close to the Kadowacki-Woods ratio $A/\gamma^2 = 1.0 \times 10^{-5}$ $\mu\Omega$ cm mol$^2$ K$^2$/J$^2$ discussed in detail for strongly correlated electronic systems.[37] This finding suggests that correlations might play some role for the present system. This $\gamma$–term, however, is considerably smaller than those reported for isostructural compounds, $Ru_3Sn_7$ ($\gamma = 19.5$ mJ/molK$^2$)[38] and $Mo_3Sb_7$ ($\gamma = 34.5$ mJ/molK$^2$).[39] As discussed below, this difference is consistent with the large number of nearly linearly dispersing electronic bands crossing the Fermi level. These bands emerge from Dirac nodes in the electronic structure of $Rh_3In_3Ge_4$.

Given that compounds belonging to the $Ir_3Ge_7$ family have been explored in the past as promising thermoelectric compounds, we evaluated the thermoelectric response of $Rh_3In_{3.4}Ge_{3.6}$. One must bear in mind that this compound displays a high degree of crystallinity, despite its off-stoichiometric composition, and its Hall response indicates a large density of carriers. Therefore, one should expect large electronic and phononic contributions to the thermal conductivity ($\kappa$) that would limit the thermoelectric figure of merit (ZT). However, a partial motivation for this study is the rich Dirac-like electronic structure observed in this compound around $\varepsilon_F$, as discussed below. In effect, it has been argued recently that the exotic electronic structures, such as topological surface states or linearly dispersing Dirac/Weyl bands, provide a fertile ground to advance different kinds of thermoelectric energy conversion schemes based on the Seebeck, magneto-Seebeck, Nernst, and anomalous Nernst effects.[40]

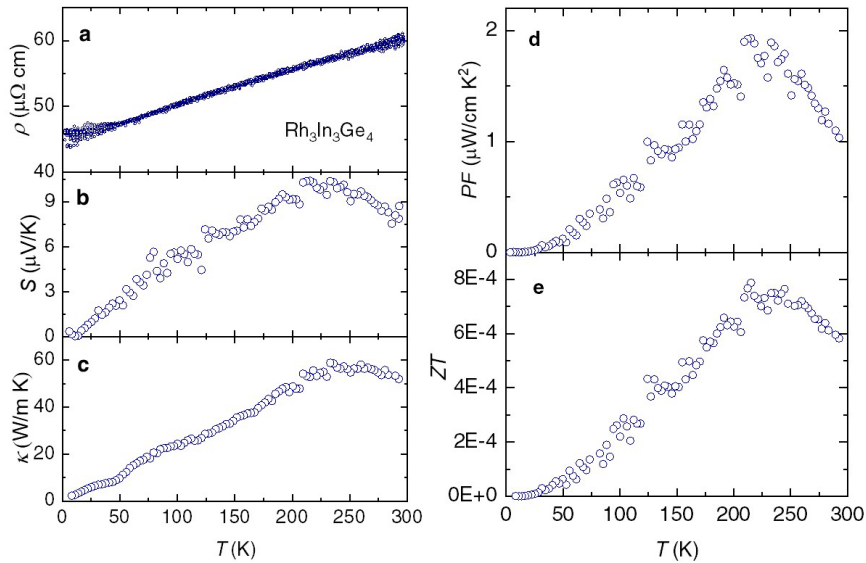

**Figure 4.** (a) $\rho$ as a function of $T$ for a $Rh_3In_3Ge_4$ single crystal used for thermoelectric measurements. (b) Seebeck coefficient $S$ as a function of $T$. (c) Thermal conductivity $\kappa$ measured on the same crystal. (d) Power factor, $PF = S^2\sigma$, where $\sigma = 1/\rho$ is the electrical conductivity as a function of $T$. (e) Resulting thermoelectric figure of merit, $ZT = S^2\sigma T/\kappa$.

An overall evaluation of a $Rh_3In_{3.4}Ge_{3.6}$ single crystal was performed to probe the temperature dependence of its thermoelectric responses, namely its resistivity $\rho$ (Figure 4a), Seebeck coefficient $S$

(Figure 4b), and the thermal conductivity $\kappa$ (Figure 4c). The value of $S$ is observed to peak around 225 K, while $\kappa$ displays a maximum around 250 K. This leads to a remarkably high value for the Power Factor, $PF = \sigma S^2 \sim 2$ μW/cm K$^2$ at $T \sim 225$ K (Figure 4d). The $PF$ is proportional to the maximum achievable thermoelectric power output of a material subjected to a certain temperature gradient, $\Delta T$. To put the extracted values in perspective, SnSe, which displays a very large thermoelectric figure of merit, $ZT = S^2\sigma T/\kappa = 2.62$ at 923 K, displays $PF$ values ranging between 2 and 10 mW/cm K$^2$ for 300 K $\leq T \leq$ 1000 K.[41] For some applications, the $PF$ may be as relevant as the $ZT$ value. In contrast to SnSe, in which both electrons and holes are claimed to contribute to electrical conduction,[41] our Hall-effect measurements indicate that holes are the main contributors to the conductivity in Rh$_3$In$_{3.4}$Ge$_{3.6}$. Ref.[42] discusses a wide range of promising hole-dominated thermoelectrics, many of which display $PF$s ranging between 1 and ~10 μW/cm K$^2$, as is the case for Rh$_3$In$_{3.4}$Ge$_{3.6}$. Nevertheless, despite the promising $PF$ displayed by this compound, its $ZT$ is rather modest due to the large phononic and electronic contributions to the thermal conductivity. Several approaches can be used to improve $ZT$, such as nanostructuring, doping, and defect engineering.[42] For example, the partial substitution of Te for Sb in isostructural Mo$_3$Sb$_7$ led to an enhancement in the thermopower by a factor ranging between 2 and 4 depending on $T$.[27] The substitution of Te for Sb was also found to enhance the ZT of Mo$_{3-x}$Fe$_x$Sb$_7$.[24]

**Electronic Band Structure.** In order to better understand the electronic properties of Rh$_3$In$_{3.4}$Ge$_{3.6}$, we performed band structure calculations on the idealized structure, Rh$_3$In$_3$Ge$_4$, through density functional theory (DFT), incorporating spin-orbit coupling while enforcing the crystallographic symmetries of the lattice (Figure 5) as encoded in the crystallographic information (CIF) file. Calculation details are provided in the Materials and Methods section.

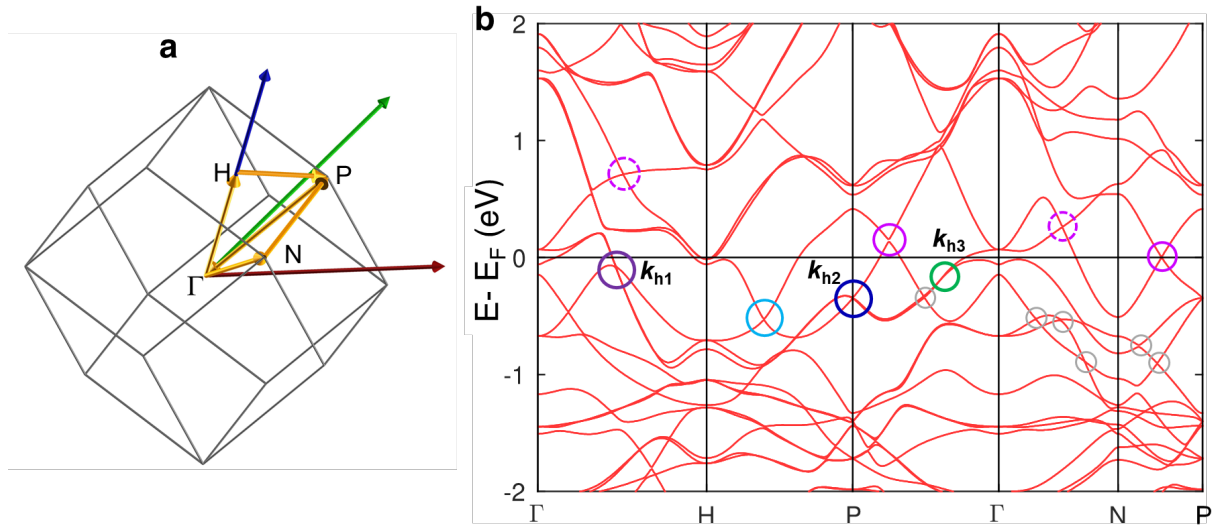

**Figure 5.** (a) First Brillouin zone (BZ) for Rh$_3$In$_3$Ge$_4$ indicating its high symmetry points and the high symmetry paths explored within its 1st BZ to evaluate its electronic band structure. (b) Electronic band structure for stoichiometric Rh$_3$In$_3$Ge$_4$ based on DFT calculations incorporating spin-orbit coupling. A series of band crossings leading to Dirac nodes are observed. Magenta circles and dashed magenta circles indicate Dirac type-I and tilted Dirac type-I nodes observed in the conduction band. As discussed in the main text, purple circle around $k_{h1}$, blue around $k_{h2}$ and green circle surrounding $k_{h3}$ indicates the location in the valence bands of a Dirac type-III, yet another type of Dirac type-III, and a Dirac type-II node, respectively. Clear blue and grey circles indicate the existence of other Dirac like crossings observed at lower energies.

The electronic band structure of $Rh_3In_3Ge_4$ reveals multiple band crossings (indicated by colored circles in Figure 5), both in the conduction and valence bands. Several of the bands crossing the Fermi level ($\varepsilon_F$) disperse linearly, as seen, for instance, between points P and Γ, forming a linear dispersion or a Dirac type-I node (Figure 1b) located at ~ 0.12 eV above $\varepsilon_F$ (indicated by a solid magenta circle). A second Dirac type-I node located precisely at $\varepsilon_F$, is also observable along the N-P path of the Brillouin zone (BZ). Dashed magenta circles indicate tilted Dirac type-I nodes located in the conduction band along directions Γ-H and Γ-N. The valence band, on the other hand, displays a plethora of linear band crossings, leading, for example, to a Dirac type-I node along the H-P direction, an apparent double Dirac structure along Γ-H (indicated by a grey circle), and to many other crossings, for instance at $k_{h1}$ (indicated by purple circle), $k_{h2}$ (dark blue circle), and $k_{h3}$ (green circle). We choose to analyze in detail the dispersion of the electronic bands around these three crossings to convey their conceptual relevance.

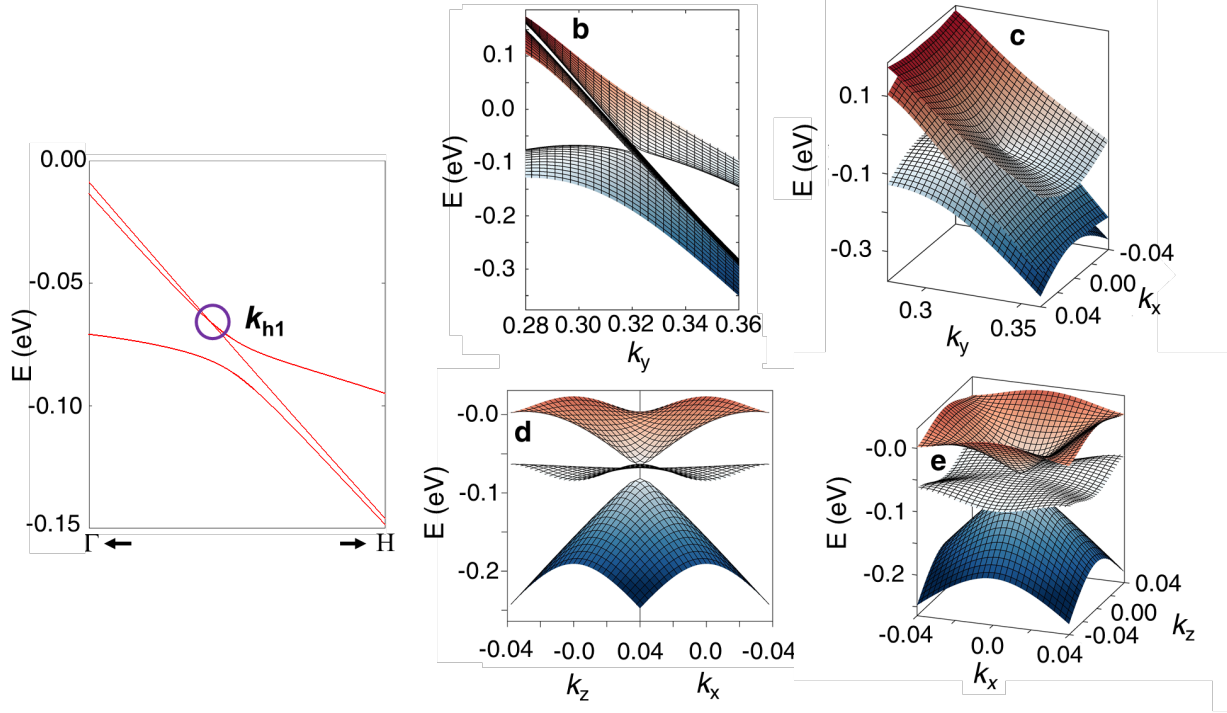

**Figure 6.** (a) Electronic band dispersion around the valence band $k_{h1}$ point. (b) Side perspective of the electronic dispersion as a function of $k_x$ and $k_y$ as depicted in (a) but with the bands plotted as 3D surfaces. (c) Distinct perspective indicating the existence of a Dirac node that intersects a flat band. (d) Side perspective of these same bands plotted as functions of $k_x$ and $k_z$. (e) Lateral three-dimensional perspective illustrating the degeneracy between the flat and linearly dispersing bands leading to a degenerate Dirac node at $k_{h1}$.

We plot the electronic dispersion (Figures 6a to 6e) around $k_{h1}$ from different perspectives in the $k$-space, where (Figures 6b, 6c, 6d and 6e) are provided to allow the reader to visualize the three-dimensional structure of this degenerate Dirac node from different perspectives. Every perspective reveals the existence of a Dirac node at the $k_{h1}$ point formed by the intersection of three valence bands located at the (0.229, -0.229, 0.229) point of the BZ, with two linearly dispersing bands intersecting an almost perfectly flat band. This is particularly clear when the bands are plotted in $k_x$ - $k_z$ plane. An intersection between a Dirac node and a flat band has been classified as a Dirac type-III node. Nevertheless, in our case, this structure does not correspond to the conventional Dirac type-III node discussed in Refs.[43-45] where it results from the tilting of a Dirac type-I cone up to a critical value where it intersects the Fermi level along a line which forms the flat band (Figure 1c). There is a some similarity between this degenerate band touching and the triply degenerate one reported for the Lieb lattice, whose flat band intersecting the Dirac node is claimed to be topological in character.[46] In our case, we see the opening of a small gap between both degenerate bands and a linearly dispersing one, which might lead to a topologically non-trivial $Z_2$ invariant.[47] Since our detailed high-resolution DFT calculations reveal just one gap at $k_{h1}$, i.e. between both degenerate bands and a linearly dispersing one, we conclude that the symmetry of the crystalline lattice must prevent gap openings among all three bands.

The structure depicted here is a unique example of a hitherto unreported Dirac type-III node in a cubic system that results from the touching of two bands, one flat and the other and a linearly dispersing, at a specific point in the Brillouin zone. Notice the proximity of this node at $k_{h1}$ to $\varepsilon_F$: it is located at just ~ -120 meV below the Fermi level. Notice that the middle flat band remains flat for $|k_x|=|k_z|$ extending beyond ± 0.04 Å$^{-1}$. Our calculations indicate that the band becomes dispersive beyond ±0.05 Å$^{-1}$ which corresponds to ~10 % of the area of the first Brillouin zone in $k_x$-$k_z$ plane (Figure S2).

Even more remarkable is the dispersion of the hole bands around the $k_{h2}$ point (Figures 7a, 7b, 7c) located at (0.25, 0.25, 0.25) within the BZ. Here, just two bands intersect at the P-point, leading to the touching (at a node) between a flat and a linearly dispersing band. Strictly speaking, this node should also be classified as Dirac type-III, although it does not result from the tilting of touching Dirac cones as originally envisioned by Refs.[43-45] (Figure 1d). To the best of our knowledge, such a structure has not been previously reported, and the unique electronic structure around this Dirac node may lead to exotic transport and topological properties of carriers. For example, strong correlations might be observed between electrons in the flat band with high effective masses. By introducing magnetic dopants, one may break the time-reversal symmetry and create the hitherto unreported Weyl type-III nodes with Berry flux in/out of these nodes around the $k_{h1}$ or $k_{h2}$ points. Similar to what is seen here, in a kagome lattice the opening of a gap between a flat band and linearly dispersing ones is claimed to lead to a $Z_2$ invariant $\neq 0$.[47]

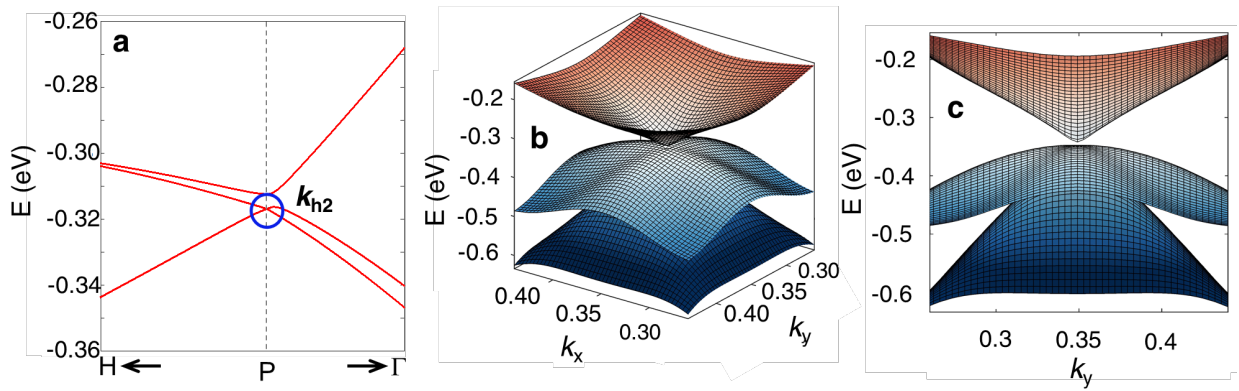

**Figure 7.** (a) Electronic band dispersion around the $k_{h2}$ point of the valence band. (b) Perspective of the electronic dispersion as depicted in (a) as a functions of $k_x$ and $k_y$ with the bands plotted as 3D surfaces. (c) Lateral perspective indicating the existence of a Dirac node that results from the intersection between a flat and a linearly dispersive band in blue) producing a particular type of Dirac type-III cone at $k_{h2}$.

The topological nature of the flat bands intersecting the Dirac nodes in Rh$_3$In$_3$Ge$_4$ remain to be clarified and will the subject of a future publication. Our calculations also indicate that the middle flat band only becomes dispersive beyond distance of ±0.025 Å$^{-1}$ with respect to $k_{h2}$ point. This indicates that this band remains flat over an area of approximately 5 % of the $k_x$ – $k_y$ plane of the first Brillouin zone.

Next we discuss the electronic structure around the $k_{h3}$ point (Figure 8) located at (0.1, 0.1, 0.1) of the BZ. Here, we observe a strongly tilted Dirac like dispersion within the $k_x$-$k_y$ plane, and, therefore, this point must be classified as a Dirac type-II node.[10, 12] Dirac and Weyl type-II dispersions are of particular conceptual relevance because the associated Dirac-like quasiparticles break Lorentz invariance, i.e. in an isotropic space the same laws of physics apply to two observers traveling at relative constant speed. To make this point clearer: in a Dirac type-I dispersion the momentum of the Landau quasiparticles (i.e. charge carriers) depends on their energy and is independent on the direction of travel. But this is not the case for tilted Dirac type-II and type-III cones, where the absolute value of the momentum becomes $k$-direction dependent.

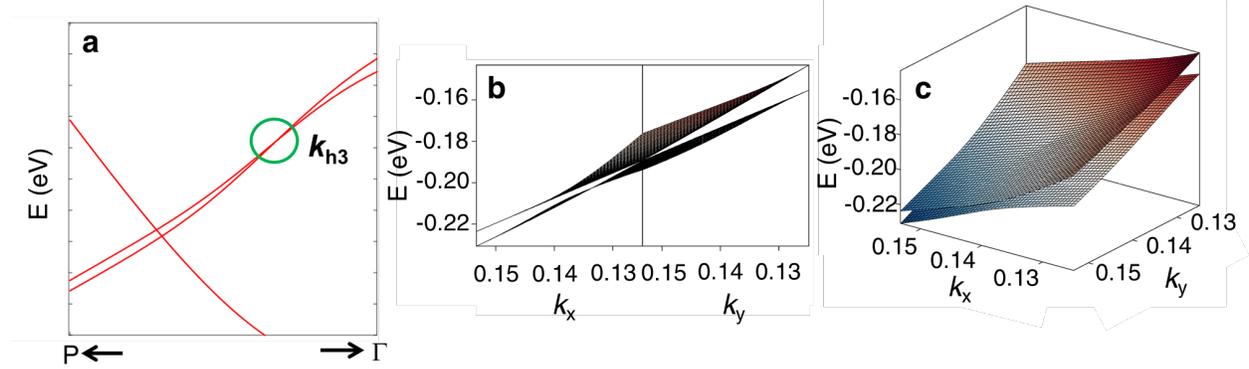

**Figure 8**. (a) Electronic band dispersion around the $k_{h3}$ point of the valence band. (b) Lateral perspective of the electronic dispersion as depicted in (a) as a functions of $k_x$ and $k_y$ with the bands plotted as 3D surfaces. (c) Distinct perspective illustrating the existence of a tilted Dirac type-II node at $k_{h3}$.

The quasiparticles associated to the originally envisioned Dirac type-III nodes are expected to exhibit extremely anisotropic effective masses μ, that is $μ \cong \infty$ for carriers traveling along the Dirac line (Figure 3c) and $μ \cong 0$ for Landau quasiparticles traveling in a direction perpendicular to it. This anisotropy in effective masses should lead to rather anisotropic transport properties. This anisotropic charge carrier behavior contrasts with the Dirac-like particles in vacuum that are described by the Dirac or Weyl equations assuming Lorentz invariance (or isotropic space). This exposes the conceptual relevance of quasiparticles associated to Dirac type-II and type-III nodes, which in the case of type-II have been predicted to yield unusual magnetotransport properties.[48]

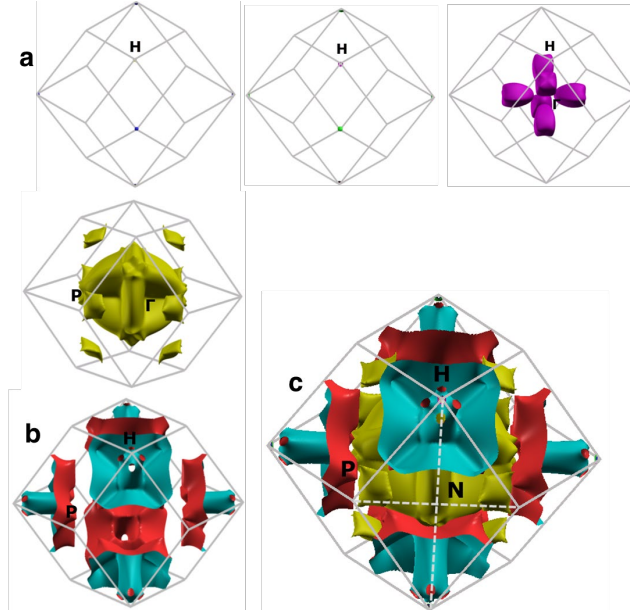

**Figure 9.** The Fermi surface of $Rh_3In_3Ge_4$ within the cubic first Brillouin zone (BZ) with the high symmetry points indicated. The Γ point is at the center of the BZ.

Finally, we discuss the geometry of the resulting Fermi surface (Figure 9) composed of electron and hole sheets. The Fermi surface reveals 12 concentric and nearly spherical electron-like pockets around the H-point, 6 hole-like pockets around the Γ-point spanning from the Γ–H direction (purple sheets), a large football-like hole sheet (in yellow) surrounding the Γ-point, and 6 large petal-like electron pockets (cyan and red sheets) spanning the H-P direction. This Fermi surface does not display a marked anisotropy that might favor magnetic or electronic orderings but exhibits a clear three-dimensional character.

## CONCLUSIONS

In summary, we synthesized, through an In-flux method, a new compound $Rh_3In_3Ge_4$ belonging to the same structure as the $Ir_3Ge_7$ family of compounds that crystallizes in the space group $Im\bar{3}m$. This compound is metallic, displays a large density of carriers according to our Hall-effect measurements, and a diamagnetic response over the entire temperature range. From a chemistry perspective, this compound represents a unique example of atomic decoration, given that the In atoms choose to occupy preferential Ge sites. Despite its large density of hole-like carriers $Rh_3In_{3.4}Ge_{3.6}$ displays a relatively large thermoelectric power factor around $T = 225$ K, albeit characterized by a small thermoelectric figure of merit. This value might improve upon proper chemical substitution or through other treatments. Remarkably, according to our calculations its electronic band structure displays a rich set of linear band crossings leading to Dirac type-I, tilted Dirac type-I, Dirac type-II and even unique examples of Dirac type-III nodes that have yet to be reported for a cubic system. All these nodes are either at, or very close in energy with respect to the Fermi level. Despite this rich Dirac like band structure, charge carriers in this compound display modest mobilities which, *via* an analogy with graphene, is attributed to scattering from the disorder inherent to its non-stoichiometric composition. In support of this hypothesis, notice that the compounds belonging to the $Ir_3Ge_7$ family tend to display similar electronic band structures while stoichiometric compounds like $Pt_3In_7$ offer residual resistivities inferior to 1 μΩ cm and carrier mobilities in excess of 3 x $10^3$ $cm^2$/Vs.[49] Another plausible scenario is the coexistence of massless linear dispersion band with massive flat bands (around the Γ-point), with the transport properties being dominated by the flat ones. It is therefore of conceptual relevance to confirm our predictions through, for example, angle resolved photoemission spectroscopy in order to unveil the existence of, for example, the Dirac type-III nodes. Given the crystallographic symmetry of this compound, the unique electronic band structure features revealed here should be degenerate and observable at other locations within its Brillouin zone. Finally, the synthesis of nearly stoichiometric crystals might allow us to unveil the true electronic and heat transport properties of this novel compound. In the Supporting Information we included a comparison between the electronic band structures of $Ru_3Sn_7$ and $Rh_3In_3Ge_4$ (Figure S3) to indicate that despite their difference in electron count, their electronic dispersion presents strong similarities displaying many band crossings, including linear ones, very near to the Fermi level. This suggests that the $Ir_3Ge_7$ family compounds is a good playground to explore the coexistance of multiple types of Dirac nodes, including Dirac type-III nodes.

## SUPPORTING INFORMATION
1. Table S1: mixed site occupancy models used for the single crystal x-ray refinement
2. Figure S1: electron energy dispersive spectra of a $Rh_3In_3Ge_4$ crystal
3. Figure S2: Electronic band dispersion around the Dirac node around $k_{h1}$ on an enlarged scale
4. Figure S3: Comparison between the electronic band structures of $Ru_3Sn_7$ and $Rh_3In_3Ge_4$

## ACKNOWLEDGMENTS
This research was supported in part by the National Science Foundation (award DMR-1905499 to M.S.). L.B. is supported by the US Department of Energy, Basic Energy Sciences program through award DE-SC0002613. S.M. acknowledges support from the FSU Provost Postdoctoral Fellowship Program. J.K.C. acknowledges the support from the DOE SCGSR Fellowship Program.  The National High Magnetic Field Laboratory acknowledges support from the US-NSF Cooperative agreement Grant number DMR-1644779 and the state of Florida. X.Q. acknowledges the support by the National Science Foundation (NSF) under award number DMR-1753054. H.W. gratefully acknowledges support from the Texas A&M University President's Excellence Fund X-Grants and T3 Program. Portions of this research were conducted with the advanced computing resources provided by Texas A&M High Performance Research Computing. The data in this manuscript can be assessed by requesting it to the correspond authors.

## MATERIALS AND METHODS

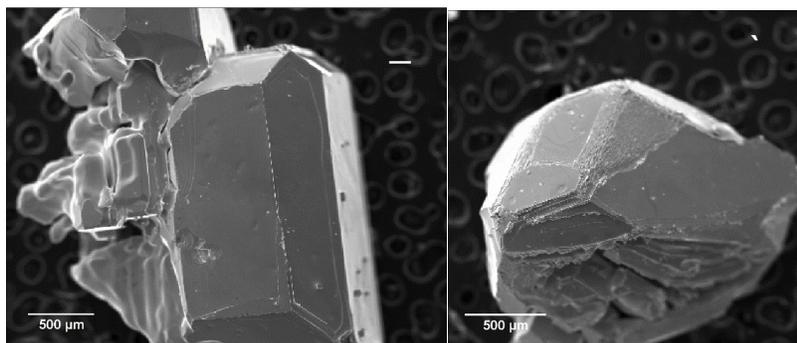

**Figure 10.** Scanning Electron Microscopy (SEM) images of RhInGe single crystals.

**Synthesis.** High-quality single crystals of $Rh_3In_3Ge_4$ were synthesized through an In-flux method: Rh (≥99.97%), In (99.99%) and Ge (99.999%) in an atomic ratio 1:20:1.3 were sealed in an evacuated quartz ampule and subsequently heated up to 1050 °C and held there for 24 h (Figure 10). Afterwards, the ampule was slowly cooled to 700 °C at a rate of 2 °C/h and subsequently centrifuged at that temperature. The as harvested single crystals were etched in HCl to remove the excess In. The outcome was large shiny crystals (Figures 9a and 9b) with dimensions up to a few millimeters. Their chemical composition was confirmed through electron dispersive x-ray spectroscopy (EDS) and single crystal x-ray diffraction (Figure S1).

**X-Ray Crystallography.** Single-crystal X-ray diffraction was performed on a Rigaku-Oxford Diffraction Synergy-S diffractometer equipped with a HyPix detector and a monochromated Mo-$K\alpha$ radiation source ($\lambda$ = 0.71073 Å). A single crystal was suspended in Parabar® oil (Hampton Research) and mounted on a cryoloop which was cooled to the desired temperature in an $N_2$ cold stream. The data set was recorded as $\omega$-scans at 0.3° step width and integrated with the CrysAlis software package.[50] A hybrid adsorption correction was applied using a numerical method based on face indexing and an empirical method based on spherical harmonics as implemented in the SCALE3 ABSPACK algorithm [51]. The space group was determined as $Im\bar{3}m$ (229). The crystal structure solution and refinement were carried out with SHELX[52] using the interface provided by Olex2.[53] The final refinement was performed with anisotropic atomic displacement parameters for all atoms. Full details of the crystal structure refinement and the final structural parameters have been deposited with the Inorganic Crystal Structure Database (ICSD). The CSD registry numbers and a brief summary of data collection and refinement parameters are provided in Table 2.

**Table 2.** Data collection and structure refinement parameters for $Rh_3In_{3.4}Ge_{3.6}$.

| Compound | $Rh_3In_{3.4}Ge_{3.6}$ |
|---|---|
| CSD Code | 1997973 |
| Temperature, K | 298 |
| $\lambda$, Å | 0.71073 |
| Space group | $Im\bar{3}m$ |
| Unit Cell $a$, Å | 8.99999(5) |
| $V$, Å$^3$ | 729.00(1) |
| Z | 4 |
| Crystal size, mm$^3$ | 0.052×0.045×0.032 |
| Crystal shape | polyhedral |
| $d_{calcd}$, g cm$^{-3}$ | 8.597 |
| $\mu$, mm$^{-1}$ | 32.033 |
| $\theta_{max}$, deg | 41.09 |
| Refl. collected ($R_{int}$) | 16488 (0.031) |
| Unique reflections | 205 |
| Parameters refined | 10 |
| $R_1$, $wR_2$ [$F > 4\sigma(F)$]$^a$ | 0.0122, 0.0269 |
| $R_1$, $wR_2$ [all data]$^a$ | 0.0122, 0.0269 |
| Goodness-of-fit$^b$ | 1.163 |
| Diff. peak and hole, e Å$^{-3}$ | 3.58, –1.46 |

$^a$ $R_1 = \Sigma||F_o|-|F_c||/\Sigma|F_o|$; $wR_2 = [\Sigma[w(F_o^2-F_c^2)^2]/\Sigma[w(F_o^2)^2]]^{1/2}$;
$^b$ Goodness-of-fit = $[\Sigma[w(F_o^2-F_c^2)^2]/(N_{obs}-N_{params})]^{1/2}$, based on all data

**Physical Properties measurements.** The samples were polished to reduce the thickness down to 0.0015 cm and gold contacts were attached with silver paint for transport measurements as seen in Figure 1e. Resistivity, Hall and heat capacity measurements were performed using a Physical Properties Measurement System (PPMS). For transport measurements a standard four-terminal method was used under several temperatures, as low as $T$ = 1.8 K and fields up to $\mu_oH$ = 9 T. Heat capacity was measured through the standard relaxation time method. Magnetization measurements were performed using a magnetic properties measurement system (MPMS) for temperatures as low as $T$ = 1.8 K and fields up to $\mu_oH$ = 9 T.

**Electronic structure of $Rh_3In_3Ge_4$**

To understand the electronic structure of $Rh_3In_3Ge_4$, we performed first-principles density-functional theory (DFT)[54] calculations using the Vienna Ab initio Simulation Package55 with the projector-augmented wave method[56] and the Perdew-Berke-Ernzerhof (PBE)[57] exchange-correlation functional within generalized-gradient approximation (GGA)[58]. We used an energy cutoff of 300 eV for the plane-wave basis and a Γ-centered $k$-point sampling of 5 × 5 × 5 for the Brillouin zone (BZ) integration. The crystal structure obtained from the x-ray experiment was used with the atomic positions relaxed by DFT. Moreover, spin-orbit coupling was included. With the wavefunctions and eigenvalues, we then constructed quasi-atomic spinor Wannier functions and tight-binding Hamiltonian by mapping Kohn-Sham wavefunctions and eigenvalues under the maximal similarity measure with respect to pseudoatomic orbitals[60, 61] using a modified WANNIER90 code[62]. As spin-orbit coupling was considered, total 184 quasi-atomic spinor Wannier functions were obtained from the projections onto the *s* and *d* pseudo-atomic orbitals of Rh, the *s* and *p* pseudo-atomic orbitals of In, and the *s* and *p* pseudo-atomic orbitals of Ge. We then used the constructed tight-binding Hamiltonian to calculate electronic band structure along high-symmetry lines and Fermi surface in a dense *k*-point grid of $100 \times 100 \times 100$.

**REFERENCES**
1. Balatsky, A. V.; Vekhter, I.; Zhu, J. X., Impurity-induced states in conventional and unconventional superconductors. *Rev. Mod. Phys.* **2006,** *78* (2), 373-433.
2. Hasan, M. Z.; Kane, C. L., Colloquium: Topological insulators. *Rev. Mod. Phys.* **2010,** *82* (4), 3045-3067.
3. Qi, X. L.; Zhang, S. C., Topological insulators and superconductors. *Rev. Mod. Phys.* **2011,** *83* (4), 1057.
4. Castro Neto, A. H.; Guinea, F.; Peres, N. M. R.; Novoselov, K. S.; Geim, A. K., The electronic properties of graphene. *Rev. Mod. Phys.* **2009,** *81* (1), 109-162.
5. Wang, Z. J.; Weng, H. M.; Wu, Q. S.; Dai, X.; Fang, Z., Three-dimensional Dirac semimetal and quantum transport in $Cd_3As_2$. *Phys. Rev. B* **2013,** *88* (12), 125427.
6. Liu, Z. K.; Jiang, J.; Zhou, B.; Wang, Z. J.; Zhang, Y.; Weng, H. M.; Prabhakaran, D.; Mo, S. K.; Peng, H.; Dudin, P.; Kim, T.; Hoesch, M.; Fang, Z.; Dai, X.; Shen, Z. X.; Feng, D. L.; Hussain, Z.; Chen, Y. L., A stable three-dimensional topological Dirac semimetal $Cd_3As_2$. *Nat. Mater.* **2014,** *13* (7), 677-681.
7. Wehling, T. O.; Black-Schaffer, A. M.; Balatsky, A. V., Dirac materials. *Adv. Phys.* **2014,** *63* (1), 1-76.
8. Zhao, Y.; Wyrick, J.; Natterer, F. D.; Rodriguez-Nieva, J. F.; Lewandowski, C.; Watanabe, K.; Taniguchi, T.; Levitov, L. S.; Zhitenev, N. B.; Stroscio, J. A., Creating and probing electron whispering-gallery modes in graphene. *Science* **2015,** *348* (6235), 672-675.
9. Peres, N. M. R., The electronic properties of graphene and its bilayer. *Vacuum* **2009,** *83* (10), 1248-1252.
10. Soluyanov, A. A.; Gresch, D.; Wang, Z. J.; Wu, Q. S.; Troyer, M.; Dai, X.; Bernevig, B. A., Type-II Weyl semimetals. *Nature* **2015,** *527* (7579), 495-498.
11. Liu, Z. K.; Zhou, B.; Zhang, Y.; Wang, Z. J.; Weng, H. M.; Prabhakaran, D.; Mo, S. K.; Shen, Z. X.; Fang, Z.; Dai, X.; Hussain, Z.; Chen, Y. L., Discovery of a Three-Dimensional Topological Dirac Semimetal, $Na_3Bi$. *Science* **2014,** *343* (6173), 864-867.
12. Noh, H. J.; Jeong, J.; Cho, E. J.; Kim, K.; Min, B. I.; Park, B. G., Experimental Realization of Type-II Dirac Fermions in a $PdTe_2$ Superconductor. *Phys. Rev. Lett.* **2017,** *119* (1), 016401
13. Amo, A.; Bloch, J., Exciton-polaritons in lattices: A non-linear photonic simulator. *C. R. Phys.* **2016,** *17* (8), 934-945.
14. Pyrialakos, G. G.; Nye, N. S.; Kantartzis, N. V.; Christodoulides, D. N., Emergence of Type-II Dirac Points in Graphynelike Photonic Lattices. *Phys. Rev. Lett.* **2017,** *119* (11), 113901.
15. Mann, C. R.; Sturges, T. J.; Weick, G.; Barnes, W. L.; Mariani, E., Manipulating type-I and type-II Dirac polaritons in cavity-embedded honeycomb metasurfaces. *Nat. Commun.* **2018,** *9*, 2194.
16. Wang, H. X.; Chen, Y.; Hang, Z. H.; Kee, H. Y.; Jiang, J. H., Type-II Dirac photons. *Npj Quantum Mater.* **2017,** *2*, 54.
17. Huang, H. Q.; Jin, Y. W.; Liu, F., Black-hole horizon in the Dirac semimetal $Zn_2In_2S_5$. *Phys. Rev. B* **2018,** *98* (12), 121110(R).

18. Liu, H.; Sun, J. T.; Song, C. C.; Huang, H. Q.; Liu, F.; Meng, S., Fermionic Analogue of High Temperature Hawking Radiation in Black Phosphorus*. *Chin. Phys. Lett.* **2020,** *37* (6), 067101.
19. Häussermann, U.; Elding-Pontén, M.; Svensson, C.; Lidin, S., Compounds with the $Ir_3Ge_7$ structure type: interpenetrating frameworks with flexible bonding properties. *Chem. Eur. J.* **1998,** *4* (6), 1007-1015.
20. Candolfi, C.; Lenoir, B.; Dauscher, A.; Tobola, J.; Clarke, S. J.; Smith, R. I., Neutron diffraction and ab initio studies of Te site preference in $Mo_3Sb_{7-x}Te_x$. *Chem. Mater.* **2008,** *20* (20), 6556-6561.
21. Yannello, V. J.; Kilduff, B. J.; Fredrickson, D. C., Isolobal analogies in intermetallics: the reversed approximation MO approach and applications to $CrGa_4$- and $Ir_3Ge_7$-type phases. *Inorg. Chem.* **2014,** *53* (5), 2730-2741.
22. Sreeraj, P.; Kurowski, D.; Hoffmann, R.-D.; Wu, Z.; Pöttgen, R., Ternary lithium stannides $Li_xT_3Sn_{7-x}$ (T = Rh, Ir). *J. Solid State Chem.* **2005,** *178* (11), 3420-3425.
23. Schlüter, M.; Häussermann, U.; Heying, B.; Pöttgen, R., Tin–magnesium substitution in $Ir_3Sn_7$ – structure and chemical bonding in $Mg_xIr_3Sn_{7-x}$ ($x$ = 0–1.67). *Journal of Solid State Chemistry* **2003,** *173* (2), 418-424.
24. Guo, Q.; Kleinke, H., The beneficial influence of tellurium on the thermoelectric properties of $Mo_{3-x}Fe_xSb_7$. *J. Solid State Chem.* **2014,** *215*, 253-259.
25. Guo, Q.; Assoud, A.; Kleinke, H., Different site occupancies in substitution variants of $Mo_3Sb_7$. *J. Solid State Chem.* **2016,** *236*, 123-129.
26. Candolfi, C.; Lenoir, B.; Leszczynski, J.; Dauscher, A.; Tobola, J.; Clarke, S. J.; Smith, R. I., Neutron diffraction, electronic band structure, and electrical resistivity of $Mo_{3-x}Ru_xSb_7$. *Inorg. Chem.* **2009,** *48* (12), 5216-5223.
27. Dashjav, E.; Szczepenowska, A.; Kleinke, H., Optimization of the thermopower of the antimonide $Mo_3Sb_7$ by a partial Sb/Te substitution. *J. Mater. Chem.* **2002,** *12* (2), 345-349.
28. Zhang, H. J.; Liu, C. X.; Qi, X. L.; Dai, X.; Fang, Z.; Zhang, S. C., Topological insulators in $Bi_2Se_3$, $Bi_2Te_3$ and $Sb_2Te_3$ with a single Dirac cone on the surface. *Nat. Phys.* **2009,** *5* (6), 438-442.
29. Hulliger, F., New $T_3b_7$ Compounds. *Nature* **1966,** *209* (5022), 500-501.
30. Swenson, D.; Sutopo; Chang, Y. A., Phase-Equilibria in the in-Rh-as System at 600-Degrees-C. *J. Alloy Compd.* **1994,** *216* (1), 67-73.
31. Sreeraj, P.; Kurowski, D.; Hoffmann, R. D.; Wu, Z. Y.; Pottgen, R., Ternary lithium stannides $Li_xT_3Sn_{7-x}$ (T = Rh, Ir). *J. Solid State Chem.* **2005,** *178* (11), 3420-3425.
32. Yan, J. Q.; McGuire, M. A.; May, A. F.; Cao, H.; Christianson, A. D.; Mandrus, D. G.; Sales, B. C., Flux growth and physical properties of $Mo_3Sb_7$ single crystals. *Phys. Rev. B* **2013,** *87* (10), 104515
33. Banszerus, L.; Schmitz, M.; Engels, S.; Dauber, J.; Oellers, M.; Haupt, F.; Watanabe, K.; Taniguchi, T.; Beschoten, B.; Stampfer, C., Ultrahigh-mobility graphene devices from chemical vapor deposition on reusable copper. *Sci. Adv.* **2015,** *1* (6), e1500222
34. Hussey, N. E., Non-generality of the Kadowaki-Woods ratio in correlated oxides. *J. Phys. Soc. Jpn.* **2005,** *74* (4), 1107-1110.
35. Debye, P., Zur Theorie der spezifischen Wärmen. *Annalen der Physik* **1912,** *344* (14), 789-839.
36. Kittel, C., Introduction to Solid State Physics. 7[th] ed.; John Wiley & Sons: New York, NY, 1996; p 157

37. Kadowaki, K.; Woods, S. B., Universal Relationship of the Resistivity and Specific-Heat in Heavy-Fermion Compounds. *Solid State Commun.* **1986,** *58* (8), 507-509.
38. Tran, V. H.; Miller, W., $Ru_3Sn_7$: Phonon Reference for Superconducting $Mo_3Sb_7$. *Acta Phys. Pol. A* **2009,** *115* (1), 83-85.
39. Tran, V. H.; Hillier, A. D.; Adroja, D. T.; Bukowski, Z., Muon-spin-rotation study of the superconducting properties of $Mo_3Sb_7$. *Phys. Rev. B* **2008,** *78* (17), 172505
40. Fu, C. G.; Sun, Y.; Felser, C., Topological thermoelectrics. *Apl. Mater.* **2020,** *8* (4), 040913.

41. Zhao, L. D.; Lo, S. H.; Zhang, Y. S.; Sun, H.; Tan, G. J.; Uher, C.; Wolverton, C.; Dravid, V. P.; Kanatzidis, M. G., Ultralow thermal conductivity and high thermoelectric figure of merit in SnSe crystals. *Nature* **2014,** *508* (7496), 373-377.
42. Wolf, M.; Hinterding, R.; Feldhoff, A., High Power Factor vs. High zT-A Review of Thermoelectric Materials for High-Temperature Application. *Entropy-Switz* **2019,** *21* (11), 1058.
43. Liu, H.; Sun, J. T.; Cheng, C.; Liu, F.; Meng, S., Photoinduced Nonequilibrium Topological States in Strained Black Phosphorus. *Phys. Rev. Lett.* **2018,** *120* (23), 237403.
44. Milicevic, M.; Montambaux, G.; Ozawa, T.; Jamadi, O.; Real, B.; Sagnes, I.; Lemaitre, A.; Le Gratiet, L.; Harouri, A.; Bloch, J.; Amo, A., Type-III and Tilted Dirac Cones Emerging from Flat Bands in Photonic Orbital Graphene. *Phys. Rev. X* **2019,** *9* (3), 031010.
45. Gong, Z. H.; Shi, X. Z.; Li, J.; Li, S.; He, C. Y.; Ouyang, T.; Zhang, C.; Tang, C.; Zhong, J. X., Theoretical prediction of low-energy Stone-Wales graphene with an intrinsic type-III Dirac cone. *Phys. Rev. B* **2020,** *101* (15), 155427.
46. Lim, L. K.; Fuchs, J. N.; Piechon, F.; Montambaux, G., Dirac points emerging from flat bands in Lieb-kagome lattices. *Phys. Rev. B* **2020,** *101* (4), 045131.
47. Kang, M. G.; Fang, S. A.; Ye, L. D.; Po, H. C.; Denlinger, J.; Jozwiak, C.; Bostwick, A.; Rotenberg, E.; Kaxiras, E.; Checkelsky, J. G.; Comin, R., Topological flat bands in frustrated kagome lattice CoSn. *Nat. Commun.* **2020,** *11* (1), 4004.
48. Yu, Z. M.; Yao, Y. G.; Yang, S. Y. A., Predicted Unusual Magnetoresponse in Type-II Weyl Semimetals. *Phys. Rev. Lett.* **2016,** *117* (7), 077202
49. Yara, T.; Kakihana, M.; Nishimura, K.; Hedo, M.; Nakama, T.; Onuki, Y.; Harima, H., Small Fermi surfaces of $PtSn_4$ and $Pt_3In_7$. *Physica B Condens. Matter* **2018,** *536*, 625-633.
50. *CrysAlis*. Oxford Diffraction Ltd.: Abingdon, England, 2006.
51. *SCALE3 ABSPACK - An Oxford Diffraction program (1.0.4,gui:1.0.3)*. Oxford Diffraction Ltd.: Abingdon, England, 2005.
52. Sheldrick, G. M., Crystal structure refinement with SHELXL. *Acta Crystallogr. Sect. C* **2015,** *71* (1), 3-8.
53. Dolomanov, O. V.; Bourhis, L. J.; Gildea, R. J.; Howard, J. A. K.; Puschmann, H., OLEX2: a complete structure solution, refinement and analysis program. *J. Appl. Cryst.* **2009,** *42* (2), 339-341.
54. Kohn, W.; Sham, L. J., Self-Consistent Equations Including Exchange and Correlation Effects. *Phys. Rev.* **1965,** *140* (4A), 1133-1138.
55. Kresse, G.; Furthmuller, J., Efficient iterative schemes for ab initio total-energy calculations using a plane-wave basis set. *Phys. Rev. B* **1996,** *54* (16), 11169-11186.
56. Blochl, P. E., Projector Augmented-Wave Method. *Phys. Rev. B* **1994,** *50* (24), 17953-17979.
57. Perdew, J. P.; Burke, K.; Ernzerhof, M., Generalized gradient approximation made simple. *Phys. Rev. Lett.* **1996,** *77* (18), 3865-3868.
58. Langreth, D. C.; Mehl, M. J., Beyond the Local-Density Approximation in Calculations of Ground-State Electronic-Properties. *Phys. Rev. B* **1983,** *28* (4), 1809-1834.
59. Dudarev, S. L.; Botton, G. A.; Savrasov, S. Y.; Humphreys, C. J.; Sutton, A. P., Electron-energy-loss spectra and the structural stability of nickel oxide: An LSDA+U study. *Phys. Rev. B* **1998,** *57* (3), 1505-1509.
60. Marzari, N.; Mostofi, A. A.; Yates, J. R.; Souza, I.; Vanderbilt, D., Maximally localized Wannier functions: Theory and applications. *Rev. Mod. Phys.* **2012,** *84* (4).
61. Qian, X. F.; Li, J.; Qi, L.; Wang, C. Z.; Chan, T. L.; Yao, Y. X.; Ho, K. M.; Yip, S., Quasiatomic orbitals for ab initio tight-binding analysis. *Phys. Rev. B* **2008,** *78* (24), 245112. .
62. Mostofi, A. A.; Yates, J. R.; Pizzi, G.; Lee, Y. S.; Souza, I.; Vanderbilt, D.; Marzari, N., An updated version of wannier90: A tool for obtaining maximally-localised Wannier functions. *Comput. Phys. Commun.* **2014,** *185* (8), 2309-2310.

**TOC Graphic**

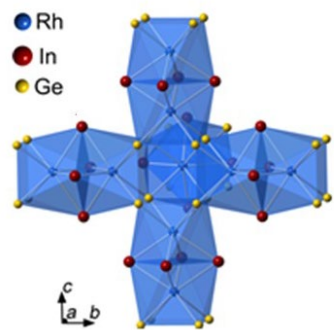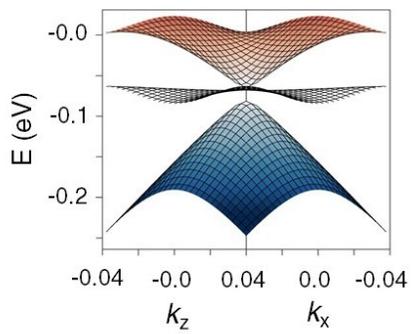

# SUPPORTING INFORMATION for manuscript titled: "Complex Dirac-like Electronic Structure in Atomic Site Ordered $Rh_3In_{3.4}Ge_{3.6}$"

Aikaterini Flessa Savvidou,[a,b] Judith K. Clark,[c] Hua Wang[d], Kaya Wei[a], Eun Sang Choi[a,b], Shirin Mozaffari[a], Xiaofeng Qian[d], Michael Shatruk,[c] Luis Balicas[a,b],*

[a] National High Magnetic Field Laboratory, 1800 E Paul Dirac Dr, Tallahassee, FL 32310, USA
[b] Department of Physics, Florida State University, 77 Chieftan Way, Tallahassee, FL 32306, USA
[c] Department of Chemistry & Biochemistry, Florida State University, 95 Chieftan Way, Tallahassee, FL 32306, USA
[d] Department of Materials Science and Engineering, Texas A&M University, College Station, TX 77843, USA

Corresponding author: balicas@magnet.fsu.edu

1. Table S1: mixed site occupancy models used for the single crystal x-ray refinement
2. Figure S1: electron energy dispersive spectra of a $Rh_3In_3Ge_4$ crystal

| In site (12$d$) occupancy | Ge site (16$f$) occupancy | Composition, Rh:In:Ge | $R$-factor, % |
|---|---|---|---|
| 100% In | 100% Ge | 3 : 3 : 4 | 2.42 |
| 91.3% In, 8.7% Ge | 16.5% In, 83.5% Ge | 3 : 3.4 : 3.6 | 1.21 |
| 86.7% In, 13.3% Ge | 86.7% Ge, 13.3% In | 3 : 3.1 : 3.9 | 1.26 |

**Table TS1**. Models used for the x-ray refinement, which assume mixed occupancy, gave a better agreement with the experimental X-ray diffraction data which is in agreement with the EDS analysis, included below.

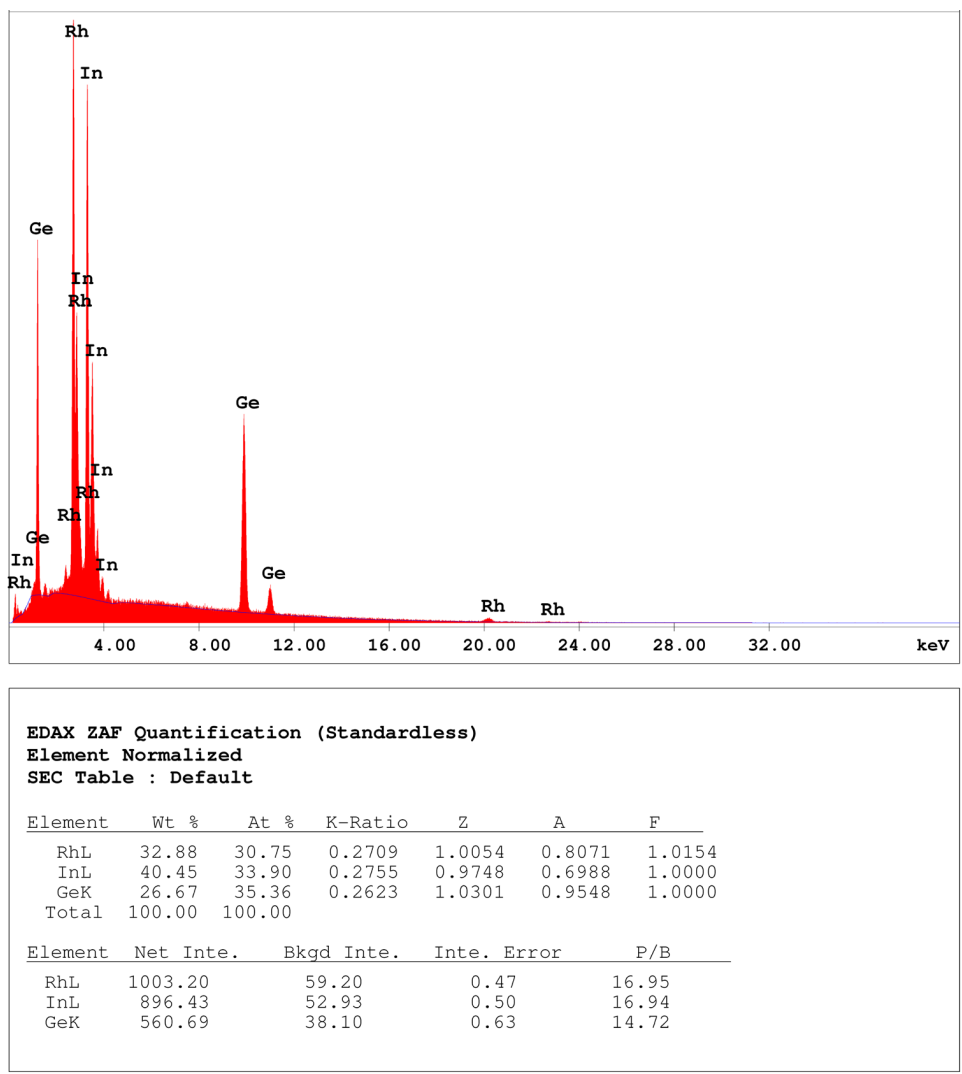

**Figure S1**. Electron energy dispersive analysis of a Rh$_3$In$_3$Ge$_4$ crystal yielding a Rh : In : Ge ratio of 3 : 3.4 : 3.6, thus fully supporting a specific mixed occupancy model yielded by the x-ray refinement.

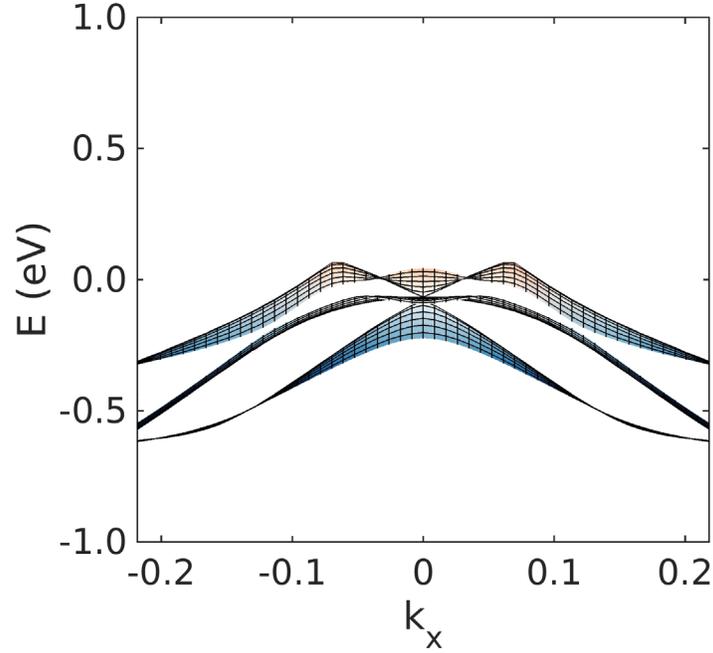

**Figure S2**. Electronic dispersion of the degenerate bands around the $k_{h1}$ point. the middle flat band only becomes dispersive beyond ± 0.05 Å$^{-1}$ away from the $k_{h1}$ Dirac node. The unit in this plot and in those within the manuscript is Å$^{-1}$. Notice that in this amplified scale the gap between the flat and the linearly dispersing bottom band is not observable.

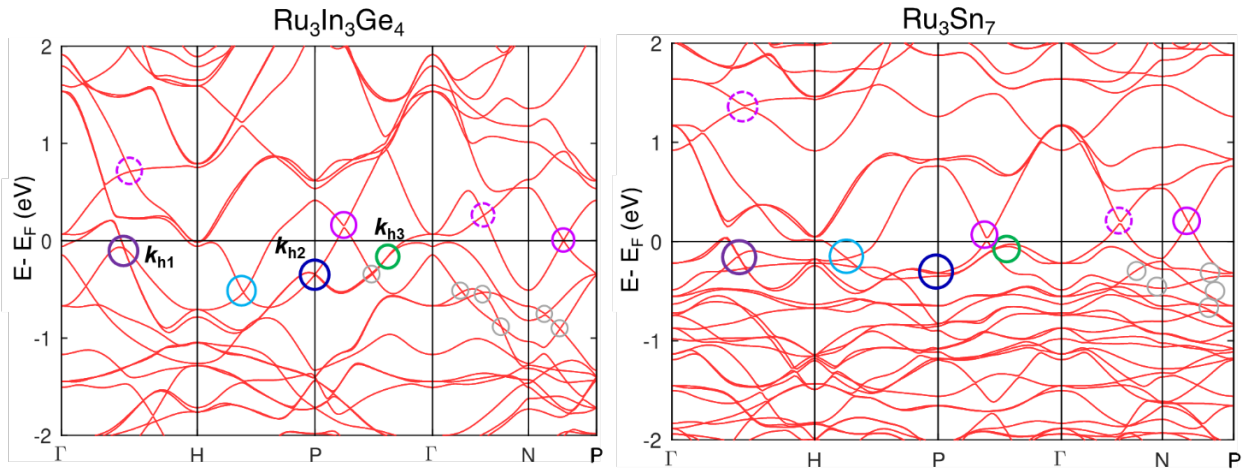

**Figure S3**. Comparison between the electronic band structures of and Rh$_3$In$_3$Ge$_4$ (left panel) and Ru$_3$Sn$_7$ (right panel). Both compounds display a large number of band crossings close to the Fermi level (indicated by circles) with some of these crossings dispersing linearly. Notice the similarities between both band structures as indicated by the colored circles, revealing akin band crossings. This suggests that the overall Ir$_3$Ge$_7$ family compounds is a good playground to explore the coexistance of multiple types of Dirac nodes.